\documentstyle[multicol,eqsecnum,prl,aps,psfig]{revtex}  

\newcommand{\fig}[3]
{\begin{figure}[htb]
\leavevmode 
\centerline{\hbox{\epsfxsize=#3cm \epsfbox{#1}}}
\caption{#2} 
\label{fig:#1} 
\end{figure}}

\begin{document}
\bibliographystyle{unsrt}

\input{epsf}
\draft

\title{Investigation of the Critical Behavior of the Critical
       Point of the Z2 Gauge Lattice}
\author{Y.~Blum$^*$, P.K.~Coyle, S.~Elitzur, E.~Rabinovici, S.~Solomon}
\address{Racah Institute of Physics, 
         Hebrew University of Jerusalem, 
         Jerusalem 91904, Israel.}
\author{H.~Rubinstein}
\address{Institutionen f\"{o}r Teoretisk Fysik, Box 803, 
S-751 08  Uppsala, Sweden 
}
\date{\today}
\maketitle

\begin{abstract}
 We investigate, through Monte-Carlo simulations, the nature of
the  second order point in a $Z_2$ (Bosonic) + $Z_2$ gauge theory in 
four dimensions. Detailed analysis of the critical exponents 
point to the Ising universality class. Relevancy to extended models and
possible Non-Gaussian behavior is discussed.
\end{abstract}

\begin{multicols}{2}
\narrowtext

In the present paper we investigate the character of the critical
point C (Fig 1) and show that the segment AC is mapped on the segment
$H=0$, $T\in [0,T_c]$ of the usual Ising model. We find that the Ising
critical exponents and the Ising universal function are reproduced to
a very high precision.

\section{Introduction}

The $Z_{2}$ gauge system has been investigated using mean-field 
method \cite{ItzyksonII}, Monte Carlo simulations \cite{Creutz,Jayaprakash} and exact 
solutions \cite {Fradkin,Rabinovici}. The objective of these studies was to find the phase diagram
from the action (equation \ref{eq:action}), and to classify the order of the phase 
transitions.

\subsection{The Model}

The action of the $Z_2$ gauge lattice can be written as
\begin{equation}
S=-\beta_{P}\sum_{\Box}U_{ij}U_{jk}U_{kl}U_{li}-\beta_{L}\sum_{(i,j)}\sigma_{i}U_{ij}\sigma_{j}
\label{eq:action}
\end{equation}
where $U_{ij}$ is the value of the link 
connecting the sites $i$ and $j$ and $\sigma_{i}$ is the value of site $i$.
The first sum is over all sets of four neighboring links (\bf{plaquettes}\rm) 
and the second sum is over all nearest neighbor sites. 
Both $U_{ij}$ and $\sigma_{i}$ can obtain the values $\pm1$ with
$U_{ij}=U_{ji}$.
The free energy is then
\begin{equation}
F=\frac{1}{N}\ln{Z}
\end{equation}
with the partition function $Z$ given by
\begin{equation} 
\label{eq:partition function}
Z=\sum_{U_{ij},\sigma_i}e^{-S}
\end{equation}
and N is the number of sites in the lattice.

From the action Eq. (\ref{eq:action})
we look at two gauge invariant observables $L$ and $P$ define 
as
\begin{eqnarray} 
P=&\langle U_{ij}U_{jk}U_{kl}U_{li}\rangle
&=-\frac{1}{4} \frac{\partial}{\partial\beta_P} 
F(\beta_L,\beta_P) \\
L=&\langle\sigma_{i}U_{ij}\sigma_{j}\rangle&=-
\frac{1}{6}\frac{\partial}{\partial\beta_L}F(\beta_L,\beta_P)
\end{eqnarray}
where the averaging is over all sites and links of the lattice. The factors $\frac{1}{4}$ 
and $\frac{1}{6}$ are the ratios between the number of sites 
to the number of links and plaquettes in a four-dimensional lattice.

In this system, $L$ and $P$ can be used as order parameters replacing the 
standard magnetization which is not gauge invariant.
They are order parameters in the sense
that they exhibit 
singularities of the bulk thermodynamics. However they lack the property of a 
magnetization in that they never vanish (except at temperature zero) or zero. 

\fig{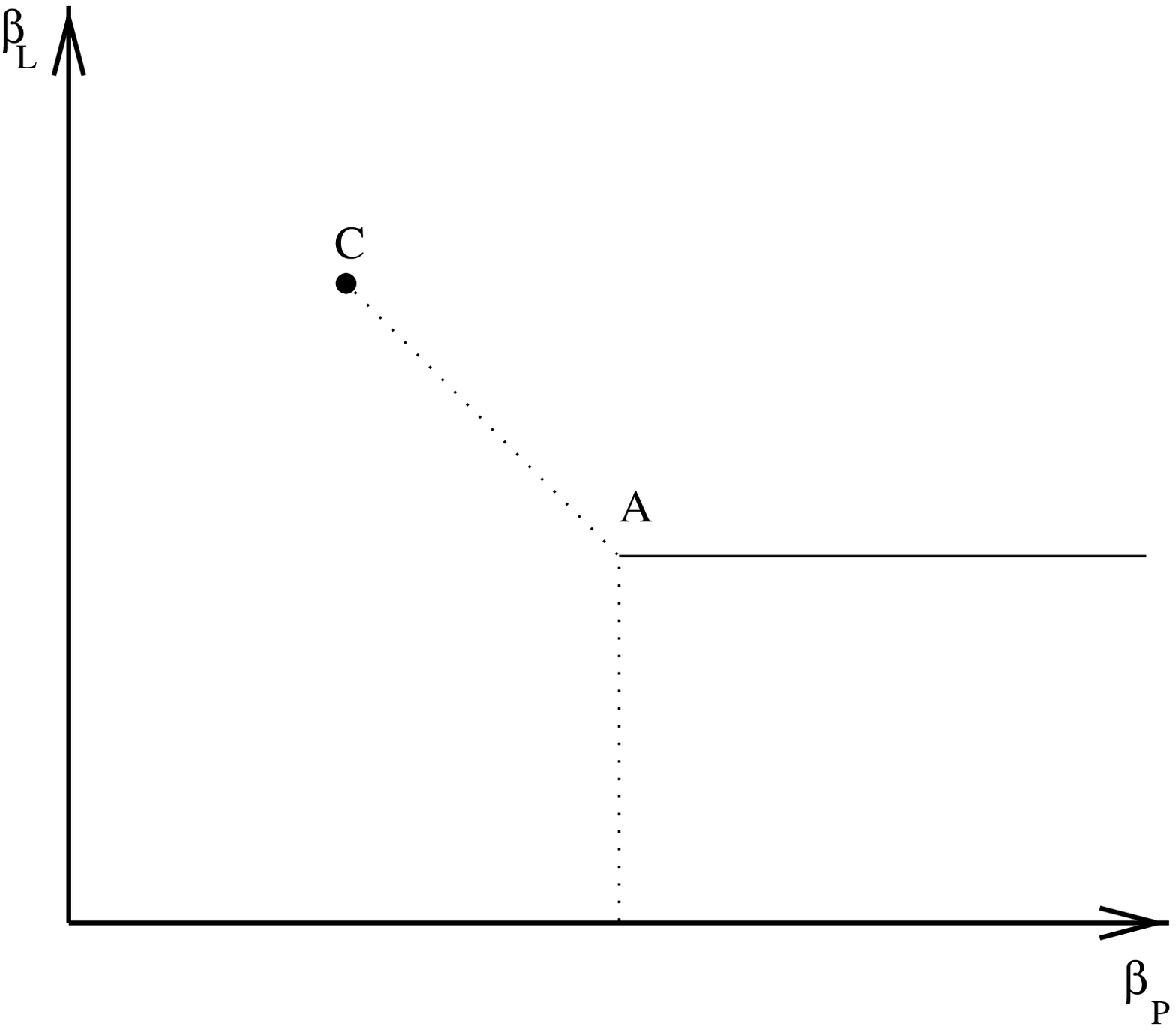}{Schematic plot of the phase diagram. The dashed lines represent 
a first order phase transition, and the continuous line represents a second 
order transition. The critical point is marked with C and the
triple point by T.}{8.0}

The phase diagram is shown schematically in Fig. \ref{fig:fig1.eps}.
The three transition lines experienced by $L$ and $P$
meet at the  \bf{triple point}\rm (A)  and split the phase space 
into three regions.
This phase space includes two familiar limits.
In the limit $\beta_{P} \to \infty$ all plaquettes
are forced to one.
 The gauge degrees of freedom therefore disappear
and the system reduces to that
of a normal four-dimension Ising model. Here $L$ experiences a second order
transition while $P$ remains smooth.
It has been shown \cite{Marcu} that the second order line is
of Ising type all the way to the triple point. 
 
 On the  $\beta_L = 0$ axis the action reduces to a pure gauge
theory and $P$ experiences a first order transition.
 This behavior continues all the way up to the
triple point A with $L$ remaining smooth \cite{Wegner}.
A line of first order in both $L$ and $P$ is then formed
which ends with the critical point C.

\section{Numerical simulation on the lattice}

We implemented a standard Monte Carlo procedure using a heat bath algorithm
to generate the configurations of the lattice. In this paper we shall use
the term \bf{iteration}\rm to represent an update of all links and sites
in the lattice.

Even using this method,  the simulations require significant computational
resources restricting us to modest lattice sizes for our investigation.
Therefore we used lattice sizes ranging from $4^4$ sites 
( which other studies have shown to give good results 
  \cite{Creutz,Jayaprakash,Marcu}) 
up to $14^4$, and at the critical point  up to $18^4$.
In order to minimize surface effects a toroidal geometry was adopted.

Each simulation consisted of measurements of the observables
along a line of points in the phase diagram.
Many such lines were made in order to cover the parameter space and
measure variations along different directions.
 At each point the
system was allowed to evolve for at least
10,000 iterations, with mean values of the observables calculated after 
discarding the first 1000.
In the vicinity of the critical point, the relaxation time 
and fluctuations 
increased dramatically requiring more iterations to give
reliable results. In this region we generated up to 100,000 iterations. 

Along the first order line $\overline{AC}$, the behavior of $L$ 
is like a magnetization with two metastable states $L^+$ and $L^-$.
However, unlike the Ising model, 
these two states are not symmetrical about the origin. 
By evaluating the mid point $L_0$ between $L^+$ and $L^-$.
we can consider $L-L_0$ which is then symmetrical about zero. 
However $L_0$ can only be found along the line of first order
where there are two stable states present.

Direct measurements of $L$ are further complicated by tunneling between
$L^+$ and $L^-$. This has the effect of blurring the first order
nature of the transition making measurements of $L_0$ 
difficult. 
In order to reduce the effects of tunneling we work with 
 $L_{rms}=\sqrt{\langle(L-L_0)^2\rangle}$.
Due to fluctuations in $L$, $L_{rms}$ will never reduce to zero even though
$\langle L-L_0 \rangle$ may become zero, however these fluctuations
decrease as the lattice size increases.

In order to measure the critical exponents, the relevant parameterization
of the phase space must be identified. 
We therefore require an
expression for the temperature $T$ and external field $H$ like directions
in terms of $\beta_P$ and $\beta_L$.

In a magnetic system, a coexistence line occurs along the temperature axis.
We therefore define the temperature axis $T$ for this system
along the line of first order $\overline{AC}$.

Along this line the reduced temperature $t$ is given by
$$
t=\frac{\beta_P}{\beta_P-\beta^c_P}
$$

The definition of the external field $H$ is 
less clear as  there is no obvious preferred direction.
In this work we used the definition used by  Br\'ezin and Drouffe 
\cite{French} in their mean field analysis of the critical exponents.

The measurements of the critical exponents were carried out in two stages.
Firstly we made a direct fit for L  to the exponents. 
This method was applied for $\beta$ and $\delta$.

Next we used finite size scaling \cite{Justin,Fisher} which gave
more reliable measurements of $\beta$ and $\delta$ and also provided
a value for $\nu$ and $\gamma$.

\section{Numerical results and discussion}
\subsubsection{Determination of $\beta$ and $\nu$}

 Figure \ref{fig:beta1.eps} 
shows how $L_{rms}$ changes with $t$ for the different lattices.  
After discarding points with small t (where $\langle L_{rms} \rangle$
differs significantly from $\langle L-L_0 \rangle$), the best fit
gives $\beta=0.41\pm0.15$. The solid line is the Ising value of $\beta=0.5$.

\fig{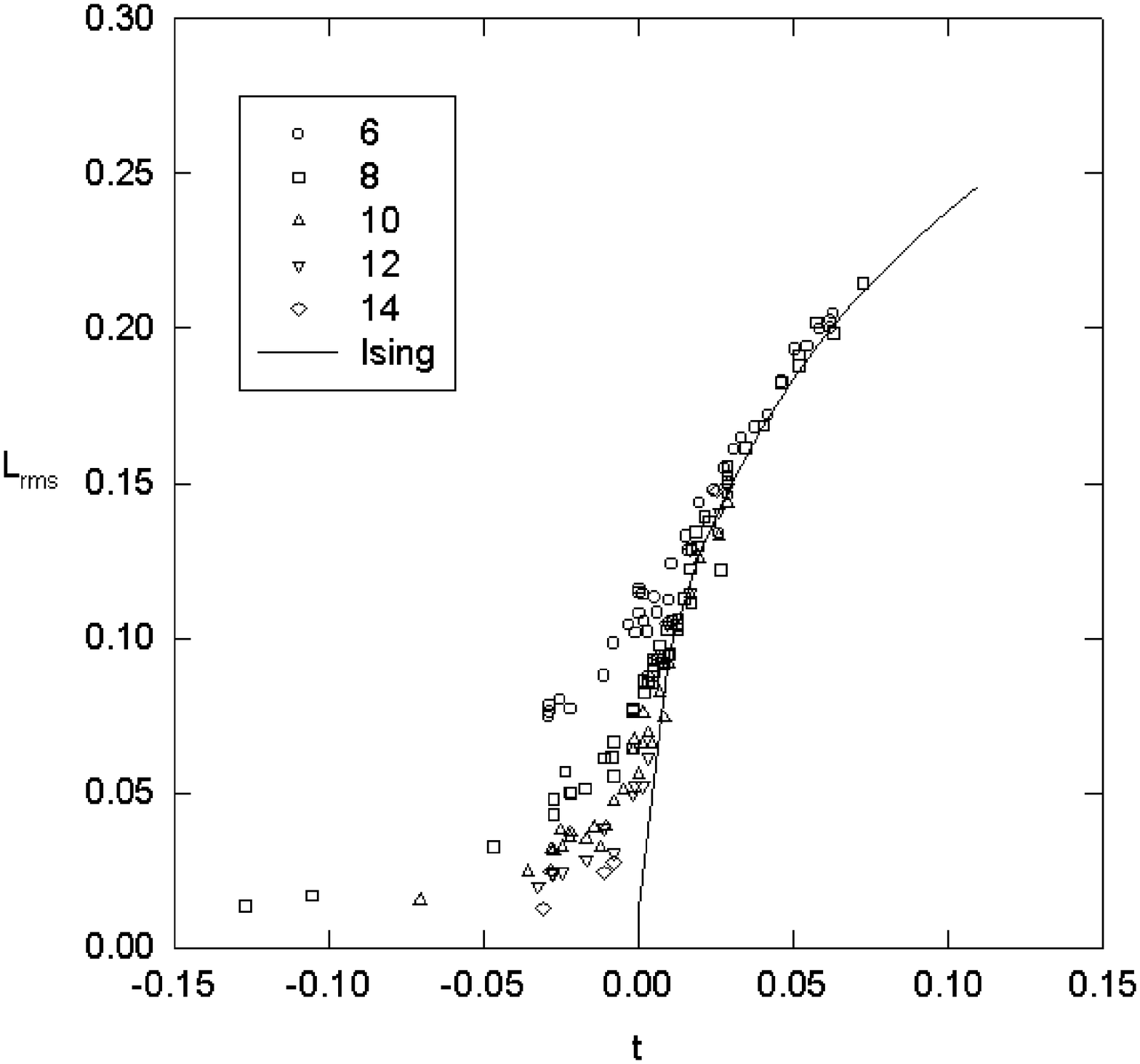}{$L_{rms}$ versus the temperature}{8.0}

A more accurate measurement of $\beta$ was obtained from finite 
size scaling. 
Figure \ref{fig:paper3.eps} shows these results and compares the data
 with simulations run in the Ising limit, $\beta_g\rightarrow \infty$ 
(solid line).
The values for the critical exponents used in this graph are those
of a four dimensional Ising model,
$\beta=0.50$ and $\nu=0.50$. 
Deviation from a single curve was observed
if $\nu$ or $\beta$ were varied by more that 0.02.

\fig{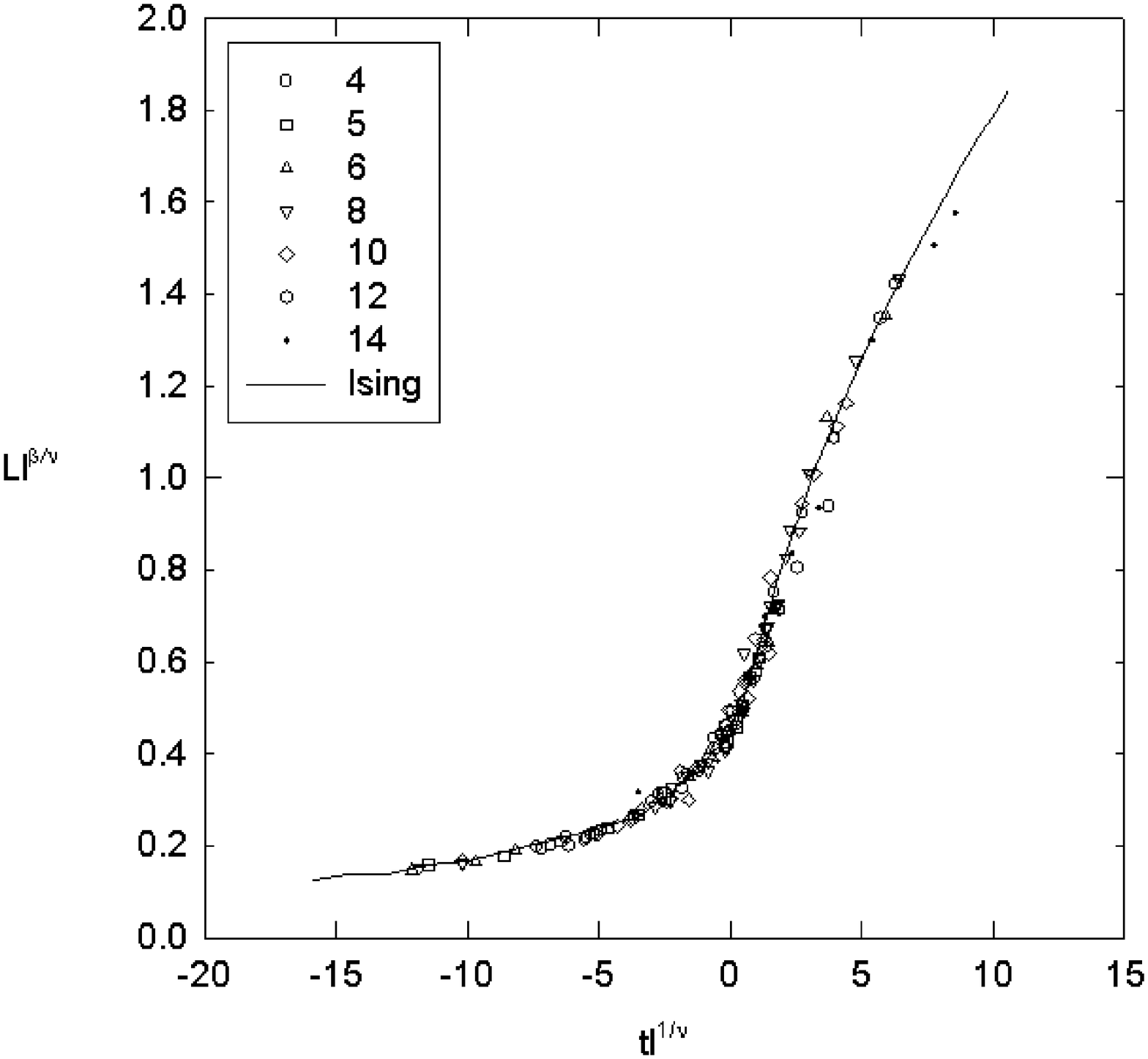}{Finite size scaling for $\beta$. $L$ is the order
  parameter $L_{rms}$, $l$ is the lattice
  size and $t$ is the reduced temperature. The solid line represents the results 
from the Ising limit.}{8.0}

\subsubsection{Determination of $\gamma$}

Direct fits with $\gamma$ were unreliable and 
depended strongly on the dimensions of the lattices 
we used.

Results from finite size scaling proved to be much more 
consistent as shown in  Fig. \ref{fig:paper4.eps}.
 Using  the value of $\nu=0.5$ found earlier
gives $\gamma=1.0\pm0.1$. 

\fig{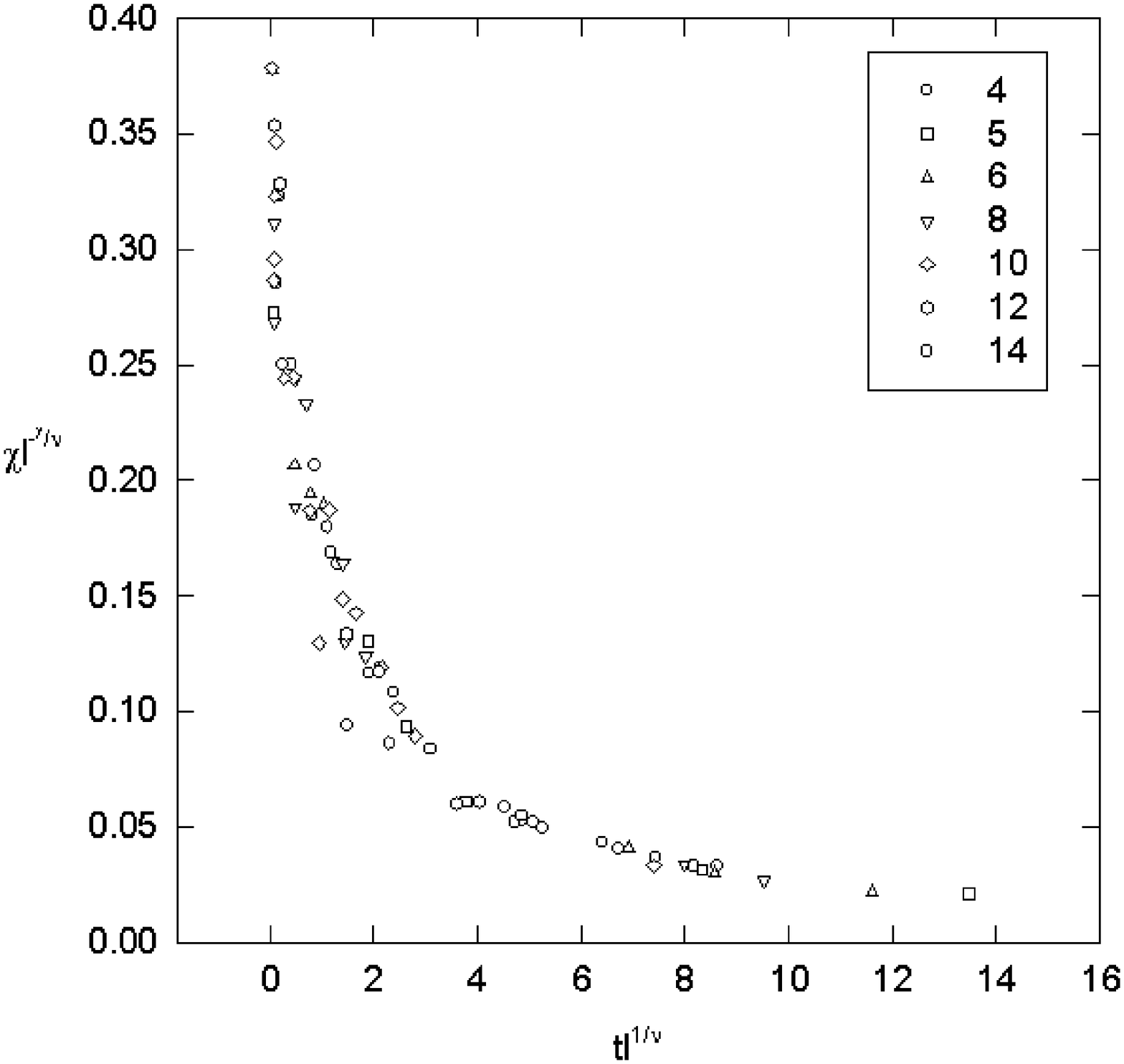}{Finite size scaling for $\gamma$. $\chi$ is the
  susceptibility, $l$ and $t$ are the lattice size and reduced
  temperature respectively.}{8.0}

\subsubsection{Determination of $\delta$}

The measurement of  $\delta$ is less straight forward.
It proved to be sensitive both to the location of the critical point
and to the lattice size (see Fig. \ref{fig:paper6.eps}).  
 The main difficulty arises from the fact that  $L$ 
is not symmetric under the operation $H \rightarrow -H$ generating different values 
either side of the temperature
axis are $\delta^+=4.4\pm$0.4 for $H>0$, and $\delta=3.0\pm0.3$ for  $H<0$. 
Measurements of lattices with different sizes showed, that the shape of
$L$ vs.$H$ becomes more symmetric as the lattice size increases. 

\fig{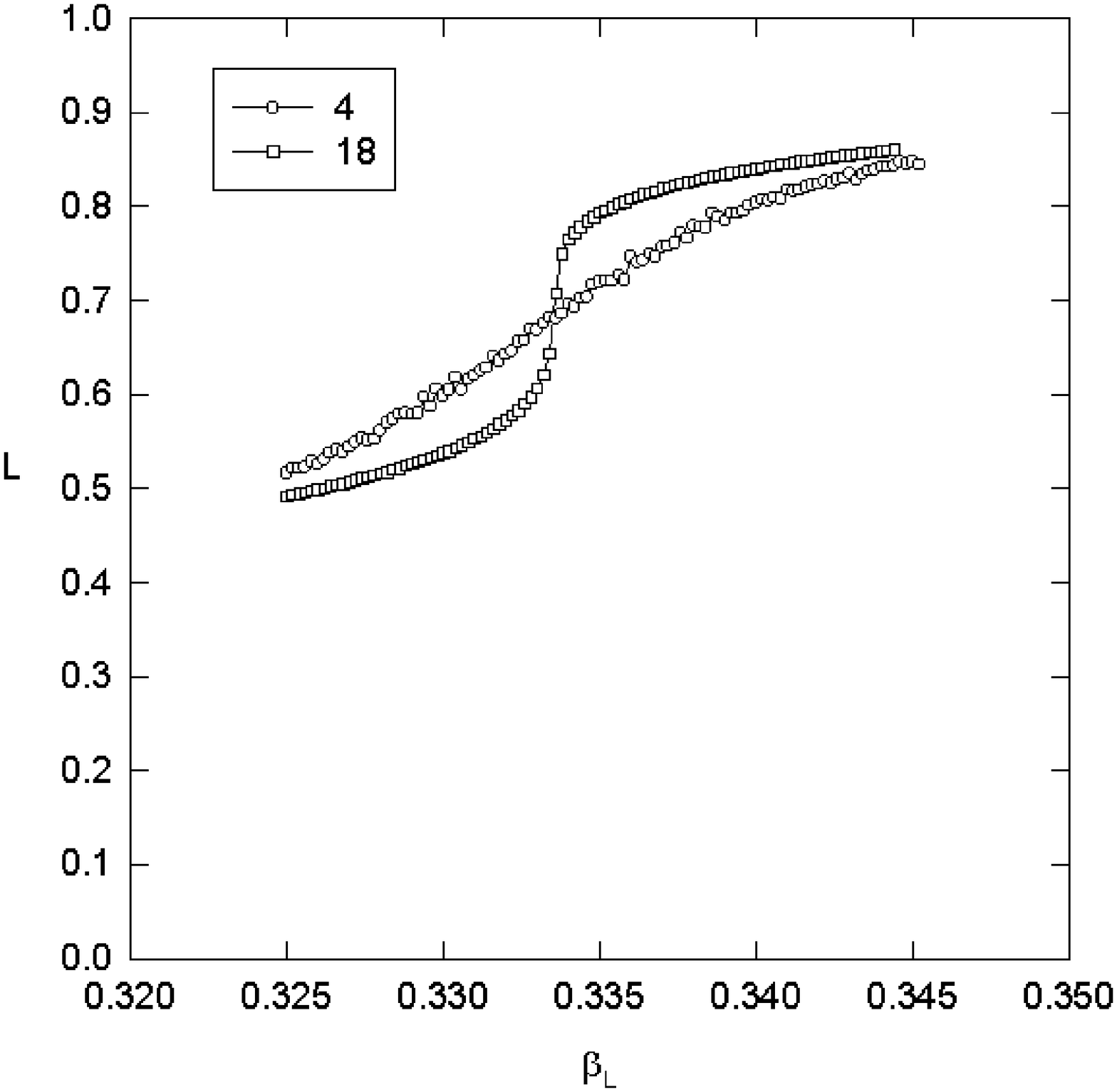}{Simulations passing through the critical point for
lattice sizes $4^4$ and $18^4$}{8.0}

The calculation of the range of $\delta$ values took place separately for
each side around the zero of $L$. The combined result is $\delta=3.7\pm$1.1.

\fig{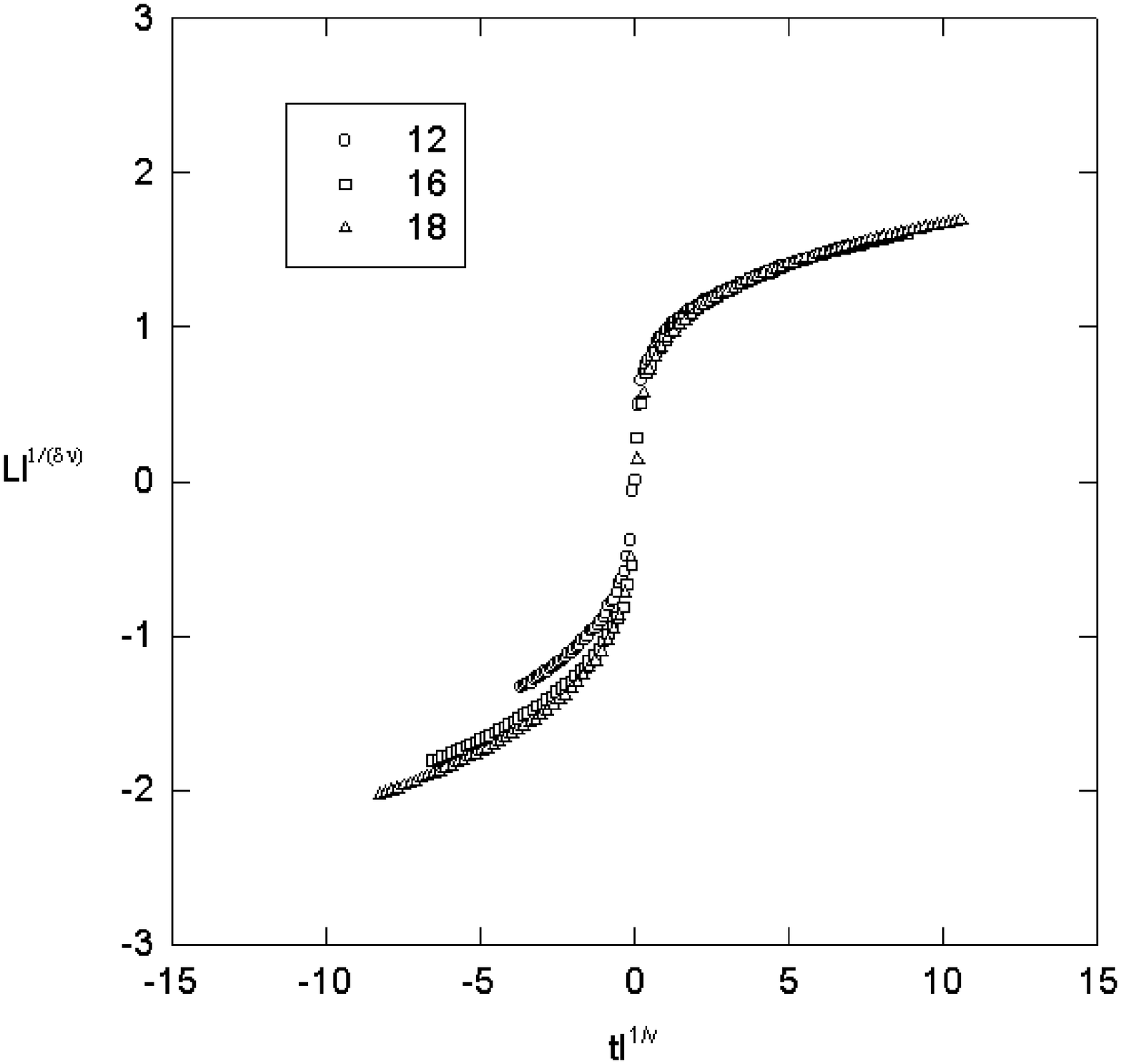}{Finite size scaling for $\delta$ at $\beta_P=0.3194$. $L$, $l$ and $t$ are defined in figure 2. The curves are not symmetrical around the coexistence
 line.}{8.0}

The asymmetry also affects finite size scaling.
 It proved impossible to collapse both 
$H>0$ and $H<0$ data onto a single curve simultaneously as shown in
Fig \ref{fig:paper5.eps}.
Treating each side independently gave $\nu=0.5$ and $\delta^+=2.5$
and $\nu=0.5$ and $\delta^-=4.7$.

\subsubsection{Universality}

The measured values of the critical exponents show that the 
critical point in this model most likely behaves like a four-dimensional
Ising model.
 The values of $\beta$ and $\nu$ fit the Ising values with very 
high accuracy, $\gamma$ has the Ising value too, but the measurement
is less accurate. 

The only critical exponent that could be  significant different from the
Ising value is $\delta$. 
We ascribe our difficulties in measuring $\delta$ to our definition
of the external field direction $\hat{H}$. 
It is not obvious from the model 
which direction is relevant and we were unable to isolate it 
from our data. 
The direction we adopted may therefore be a combination of $T$ and
$H$, thus our $\delta$ measurement is probably contaminated with effects
from $\beta$.

\section{Summary}

This work is a numerically investigation of the behavior of the critical point 
in the $Z_2$ gauge lattice. The critical exponents
were measured using Monte Carlo simulations, and 
were calculated directly as well as through finite size scaling. Other 
measurements were made in the Ising limit of the system. 

The calculated exponents had the four-dimensional Ising model values.
The numerical measurements gave the same values as the theoretically calculated 
exponents of $\beta$ and $\delta$. Two other critical exponents, $\nu$ and 
$\gamma$, that were measured, gave also the Ising values. 
The finite size scaling for $\beta$ created the same shape of curve as the one 
obtained from Ising data.
The present high precision runs set a high precision standard for the
estimation of the character of other interesting critical points,
notably the vortex/monopole condensation points in mixed action gauge
theories and analogue points in supersymmetric theories.

We would like to thank Professor G. Parisi for the helpful discussions.
This work is supported in part by BSF -- American-Israel Bi-National
Science Foundation, and by the Israel Science Foundation founded by the
Israel Academy of Sciences and Humanities -- Centers of Excellence Program.

\end{multicols}

\begin{references}

\bibitem[*]{address} Current address:
School of Physics and Astronomy, Raymond and
Beverly Sackler Facullty of Exact Sciences, Tel Aviv University, Tel
Aviv 69978, Israel.

\bibitem{ItzyksonII}
R.Balian, J. M. Drouffe, and C. Itzykson, 
Phys. Rev. D\bf{11}\rm, 2098 (1975).

\bibitem{Creutz}
M.~Creuts,
Phys. Rev. D\bf{21} \rm (1980), 1006.

\bibitem{Jayaprakash}
J.~D.~Stack G.~A.~Jongeward and C.~Jayaprakash, 
Phys. Rev. D\bf{21} \rm (1980), 3360.

\bibitem{Fradkin}
E.~Fradkin and S.~H. Shenker,
Phys. Rev. D\bf{19} \rm (1979), 3682.

\bibitem{Rabinovici}
T. Banks, E. Rabinovici, Nucl. Phys. B\bf{160} \rm(1979), 349.

\bibitem{Marcu}
M.~Marcu T.~Filk and K.~Fredenhagen,
Phys. Lett. B\bf{169} \rm (1986), 405.

\bibitem{Wegner}
F.~Wegner,
J. Math. Phys. \bf{12} \rm (1971), 2259.

\bibitem{French}
E.~Br\'ezin and J.~M. Drouffe,
Nucl. Phys. B\bf{200} \rm (1982), 93.

\bibitem{Justin}
E.~Br\'ezin and J.~Zinn-Justin,
Nucl. Phys. B\bf{257} \rm (1985), 867.

\bibitem{Fisher}
M.~E. Fisher and M.~N. Barber,
Phys. Rev. Lett. \bf{28} \rm (1972), 1516.

\end{references}
\end{document}